\documentclass[10pt, conference, compsocconf]{IEEEtran}

\ifCLASSINFOpdf
\else
\fi

\usepackage{graphicx}
\usepackage{hyperref}
\usepackage[numbers]{natbib}
\usepackage{amsmath, amsthm, amssymb, setspace, mathrsfs, color, verbatim, graphicx, amsfonts, bm, rotating, lscape, multicol, multirow,bbm, psfrag, epsf, enumerate} %natbib
\usepackage{soul}
\usepackage[section]{placeins}
\usepackage{float}
\usepackage{subcaption}
\usepackage{longtable}

\newcommand{\widthcoef}{0.48}

\begin{document}

\title{Mixed Data and Classification of Transit Stops}

\author{\IEEEauthorblockN{Laura L. Tupper\IEEEauthorrefmark{1},
David S. Matteson\IEEEauthorrefmark{2}, and
John C. Handley\IEEEauthorrefmark{3}}
\IEEEauthorblockA{\IEEEauthorrefmark{1}Department of Mathematics and Statistics\\
Williams College,
Williamstown, Massachusetts 01267\\ Email: llt1@williams.edu}
\IEEEauthorblockA{\IEEEauthorrefmark{2}Department of Statistical Science\\
Cornell University,
Ithaca, NY 14853\\ Email: matteson@cornell.edu}
\IEEEauthorblockA{\IEEEauthorrefmark{3}PARC, a Xerox Company, 800 Phillips Road, MS 128-27E, Webster, NY 14580\\
Email: John.Handley@parc.com}}

\maketitle

\begin{abstract}
An analysis of the characteristics and behavior of individual bus stops can reveal clusters of similar stops, which can be of use in making routing and scheduling decisions, as well as determining what facilities to provide at each stop. This paper provides an exploratory analysis, including several possible clustering results, of a dataset provided by the Regional Transit Service of Rochester, NY. The dataset describes ridership on public buses, recording the time, location, and number of entering and exiting passengers each time a bus stops. A description of the overall behavior of bus ridership is followed by a stop-level analysis. We compare multiple measures of stop similarity, based on location, route information, and ridership volume over time.

\end{abstract}


\IEEEpeerreviewmaketitle

%%%%%%%%
\section{Introduction}
\label{sec:xer_intro}
%%%%%%%%

City and regional planners, investors, and transit authorities have long sought to understand how people use public transit, and how they can be induced to do so more extensively and efficiently---a question that becomes still more important as planners grapple with static or shrinking budgets \cite{chiang}. Effective policies and decisions about scheduling, routing, and the facilities available to riders can encourage ridership while making optimal use of budget, fuel, and other resources; and this decision-making process benefits from concise, interpretable characterizations of both system-wide and local behavior. In large transit systems, it is particularly desirable to find commonalities and patterns between different locations or stops within the overall system.

There is extensive research on the drivers of transit ridership, surveys of which can be found in \cite{taylor09} and \cite{taylor13}. They conclude that ``external factors" beyond the control of transit authorities, relating to the site, demographics, and built environment of the city, are the dominant factor in transit ridership; but they also note that factors related to the transit service itself, like reliability, schedule convenience, and pricing schemes, play a limited but significant role.

These studies take a system-wide view of ridership, but system-wide performance is ultimately an aggregate of performance at local areas. Several studies, such as \cite{cardozo} and \cite{gutierrez}, attempt to model or forecast ridership at individual transit stops or stations, with \cite{cardozo} using geographically weighted regression to account for the gradually decreasing influence of the local area as the distance from the stop increases. Further complexity arises when ridership behavior is considered over time as well as over location; as \cite{goodchild} demonstrate with their study of time-location diaries in Halifax, both individuals and the distribution of people in an urban area vary greatly over the course of a day. A study of taxi ridership in \cite{liu} reveals similar shifting patterns in travel, associated with spatial factors of land use as well as time. In a transit-specific context, \cite{verbas} examine the elasticity of demand at a given transit stop in response to changes in service, noting that this depends both on location and on time. 

\cite{goodchild} attempt to reduce the dimension of the complex space of individual behavior through factor analysis. To a transit planner, however, individual movements can be viewed through the lens of rider interaction with transit stops. The problem then becomes the efficient characterization of transit stops, whether in terms of their use, their network characteristics, or their surrounding environment. Ideally, stops could be classified into a limited number of types, allowing reasonable performance comparisons across stations and different policy decisions for each station type, as discussed in \cite{zemp}.

There have been several approaches to the characterization, and classification, of transit stops. One influential idea is the ``node-place" model described in \cite{bertolini} in an analysis of Dutch transit stations. ``Node" factors relate to the stop's positioning in the network: its connection to other stops or other transit modes, or the range of locations, landmarks, or jobs accessible from the stop. Some measures of accessibility are described in \cite{fu} and \cite{boisjoly}. ``Place" factors, in contrast, describe the local environment of the stop: the range of activities and opportunities available at that location. \cite{bertolini} concludes that the balance between these factors is an important consideration when determining institutional investment, whether in developing the area of a station or improving its connectivity to the rest of the network. This idea of investment and planning decisions as motivation for classification is echoed by \cite{higgins}, who examine current or possible transit station sites for their potential for transit-oriented development.

The node-place model is extended in several other attempts at classification of transit stations. \cite{reusser} use node and place features of Swiss rail stations not only to characterize the overall tendencies of stations within the network, but to cluster stations into distinct and potentially interpretable types. \cite{zemp} further refine the model, again on Swiss rail stations, emphasizing stations' ``system context" and examining the associations between clusters, geographic locations, and usage data. \cite{vale} adds a measurement of ``urban design" to the classification features, based on the local area's suitability for pedestrian use.

Other researchers have drawn on usage patterns to classify stations. \cite{chen} base a clustering of New York City subway stations on diurnal patterns of passenger volume; instead of working directly with the time series, they identify features such as the starting times and durations of peak periods and cluster on these features. \cite{kim} perform a similar analysis of Seoul subway stations, starting with hourly passenger volume and using PCA to reduce dimension before clustering. 

In planning-oriented discussions like \cite{higgins} and \cite{zemp}, actual usage of a transit stop is considered an output rather than an input, and certainly it is not possible to ignore the fact that transit supply and demand are mutually influential. Nonetheless, many transit-related decisions can be based on observed usage. It is also worth noting that a time-varying characterization of usage---rather than a single measurement of passenger volume, of which \cite{zemp} are highly dismissive---may reflect the purposes and types of travel occurring at a stop, not only its popularity.

This paper explores characterization and classification problems using ridership data from the Rochester, NY bus system, both at the aggregate level and broken down by time and location. A description of the dataset, with an exploration of temporal patterns at a system level and of ridership distribution across locations, is given in Section \ref{sec:xer_data}. Section \ref{sec:xer_explo} includes a stop-level analysis of ridership, using several methods to measure similarity and dissimilarity between sites, and presenting an example of clustering sites. Finally, Section \ref{sec:xer_conclu} gives our conclusions and a discussion of future work.

%%%%%%%%
\section{Data and exploratory analysis}
\label{sec:xer_data}
%%%%%%%%

The current dataset is provided by the Regional Transit Service (RTS) of Rochester, NY. Rochester is a city of about 200,000 residents, with just over one million residents in the metropolitan area; although it is connected to inter-city rail, its local public transit consists entirely of buses and paratransit shuttle services. The current dataset focuses on bus routes within the city of Rochester itself. It is important to note that the Rochester bus system follows a ``hub" model, in which most routes lead to or from a single point, the downtown Rochester Transit Center (or, for traffic flow reasons, a location on one of the surrounding streets).

The data are gathered from buses equipped with time measurement and automated passenger count (APC) equipment. As the bus proceeds along its route, it records location and time; when the bus reaches a stop and opens its doors, the APC equipment records the number of people entering (``boarding") or exiting (``alighting") from the bus. An entry in the dataset, then, consists of one such door-opening event, with its location, time, and number of boardings or alightings, along with information about the bus route and cumulative passenger load. An entry may still be generated if no one enters or exits the bus.

Although the dataset is massive, covering nearly a year of data from dozens of RTS bus routes, this exploratory analysis focuses on a smaller subset. The route of interest here is route 39/39X, named ``Bay/Webster" for the main streets along which it travels. This route is a useful representative since it runs for most of the day at a respectable frequency (usually every hour or half-hour, with additional service during morning peak times), passes through distinct areas of downtown Rochester including the city center and more residential areas, and is relatively short, making visualization of the data more manageable. 

\begin{figure}[H]
\centering
\includegraphics[width=\widthcoef \textwidth]{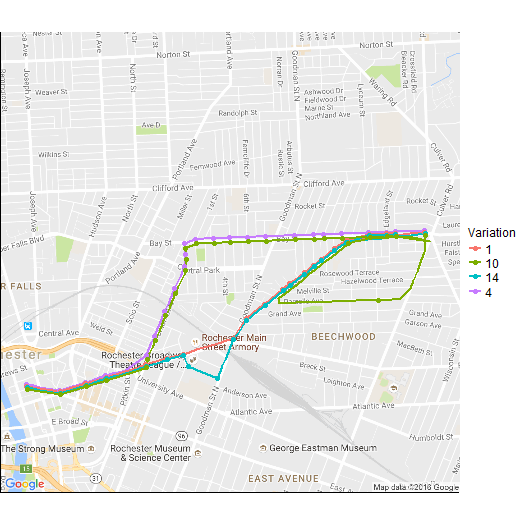}
\caption{Map of route 39 inbound variations. Stops marked have at least 50 boardings during the study period (see below); low-volume stops are not marked.}
\label{39I_vmap}
\end{figure}

Route 39 trips are classified as ``Inbound," originating in northeast Rochester and terminating at the Transit Center, or as ``Outbound" for the reverse journey. As with most RTS routes, buses on route 39 follow one of several route variations depending on time of day. This practice presents a challenge to the analyst, since not all stops appear on all routes, and their ridership may also depend on which other stops appear on the route. The inbound variations of route 39 are shown in Figure \ref{39I_vmap}, with stops of significant ridership volume marked.

%% varn map

The dates of interest are also restricted, since RTS updates its routing and scheduling regularly based on usage and seasonal changes such as the beginning and end of the school year. (RTS provides busing for many Rochester high school students, so its ridership profile is strongly affected by the academic calendar.) This analysis uses data from schedule 81, which covers weekday service from January 26 to April 3 of 2015. Since this period includes no major holidays or school breaks, it provides fairly consistent data. To account for any seasonal trends, and for covariates such as weather and local events that may affect ridership on a week-to-week basis, the data can be aggregated across the nine weeks of schedule 81.

%% ridership by DOW
\begin{figure}[H]
\centering
\begin{subfigure}{.21\textwidth}
\includegraphics[width= \textwidth]{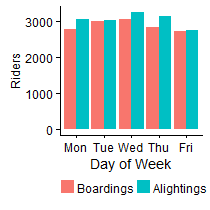}
\caption{Total riders.}
\end{subfigure} %\quad
\begin{subfigure}{.21\textwidth}
\includegraphics[width= \textwidth]{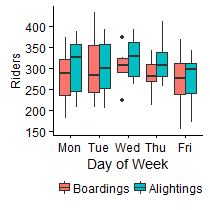}
\caption{Boxplot of variation across weeks.}
\end{subfigure}
\caption{Boardings and alightings by day of week, inbound trips on route 39.}
\label{39I_DOW}
\end{figure}

As seen in Figure \ref{39I_DOW}, ridership tends to be higher midweek and lower on Mondays and Fridays. The plot of total volume also reveals a measurement problem: in this example, more riders leave the bus than ever boarded it. This discrepancy is due to measurement error when the APC equipment records the number of riders entering or exiting the bus during a doors-open event. This problem is especially pronounced at high-volume stops, where riders may be crowded closely together as they enter or exit, confusing optical measurement. It cannot be remedied in this analysis; but in future it may be desirable to complement APC measurements with farebox data or in-person observation. 

\begin{figure}[h]
\centering
\begin{subfigure}{.45\textwidth}
\includegraphics[width= \textwidth]{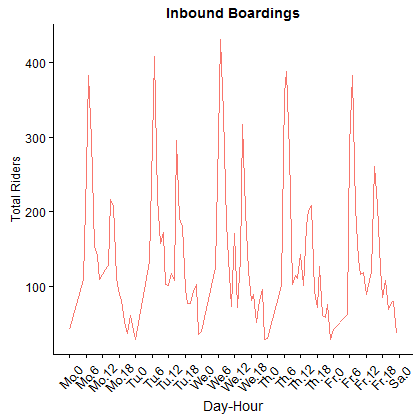}
\caption{By day of week and hour of day.}
\label{39I_HDOW_Bo}
\end{subfigure} \quad
\begin{subfigure}{.45\textwidth}
\includegraphics[width= \textwidth]{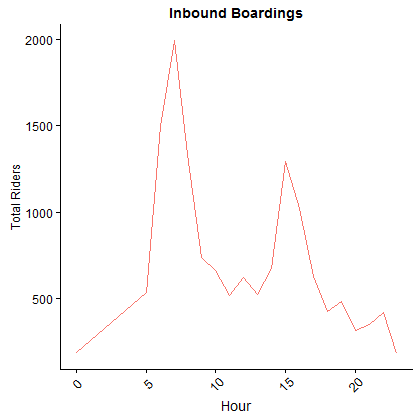}
\caption{By hour of day.}
\label{39I_H_Bo}
\end{subfigure}
\caption{Aggregated boardings for inbound trips on route 39.}
\end{figure}

Next, boarding and alighting data can be examined on an hourly basis. Despite the difference in mean levels over the course of the week, there is a fairly consistent diurnal pattern for all five days, emphasized in Figure \ref{39I_H_Bo} by aggregating the data across days of the week.

%% ridership by HDOW

This picture of overall route behavior may be expanded to a stop-level analysis. A challenge arises here in the sheer number of stops, many of which are little used. Compared to rail and subway systems, local bus routes have a very high number of stops. These stops may be easily moved with schedule or routing changes, and a given trip does not necessarily have to stop at each one; accordingly, a location need not have a particularly high ridership volume to merit a bus stop. Many stops report only a few boardings in a day, or none at all. At the other extreme, there are a few stops with extremely high ridership volume: the Transit Center is the prime example, but a few high schools and employment centers also fall into this category.

Route 39 serves 96 different stops (including paired stops for inbound and outbound use, usually located across the street from one another), and the volume of riders at each does follow this skewed distribution. In particular, the Transit Center has over 10000 boardings during this period, while the median is 70.5. A log transformation clarifies the boarding counts somewhat; the Transit Center outlier is still clear, as well as a group of stops with very low total boardings (less than 10 throughout the period).

The ridership data from these little-used stops are difficult to use in analysis; it is unreliable to estimate a diurnal curve for a stop that sees one boarding every few days. The analysis below, therefore, considers only stops with a minimum overall demand, requiring that the total number of boardings throughout the study period be above 50 riders. 41 inbound stops meet this criterion, but only 11 outbound stops. This discrepancy in distribution reflects the orientation of trips on route 39 toward the city center: riders board from many locations to travel into the city, but outbound traffic originates predominantly from the Transit Center and a few surrounding stops. Alightings display the reverse of this pattern: the destination of most inbound trips is the Transit Center, while outbound riders exit the bus at a wider range of stops.

Stops do not, however, merely differ in their overall volume of ridership; different stops display different use profiles over the course of a day. The next section examines these diurnal patterns, and which stops behave similarly in this sense.

%%%%%%%%
\section{Distance and classification}
\label{sec:xer_explo}
%%%%%%%%

\begin{figure}[h]
\centering
\begin{subfigure}{.45\textwidth}
\includegraphics[width= \textwidth]{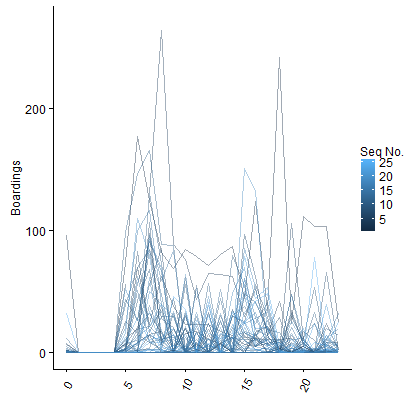}
\caption{Absolute boarding counts.}
\label{ni39IBo_h}
\end{subfigure}\quad
\begin{subfigure}{.45\textwidth}
\includegraphics[width= \textwidth]{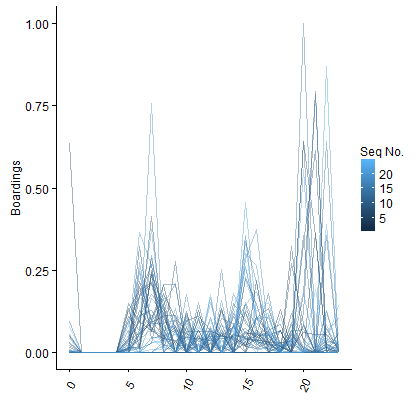}
\caption{Proportions of total boardings.}
\label{ni39IBo_hP}
\end{subfigure}
\caption{Stop-level patterns of boardings, by hour of day. Color corresponds to the stop's position along the route, with lighter-colored stops near the end of the route.}
\label{ni39IBos}
\end{figure}

Breaking down ridership to the stop level shows results in agreement with the route-wide results. There is some fluctuation in level between days of the week, but a recognizable diurnal pattern appears each day, with two distinct peaks. This behavior is more clearly seen when days of the week are aggregated to obtain a single diurnal pattern for each stop, as in Figure \ref{ni39IBos}. A few stops dominate the boardings; reasonably, for inbound trips, the most-used stops are at the beginning of the route and near the end of the route, when the bus is passing through the dense downtown area.

Here, again, the distinctions between stops depend heavily on overall level; the few high-volume stops stand out, while other stops are difficult to distinguish. To consider stops' pattern of ridership over the course of the day, rather than the overall volume, hourly boarding counts are converted to proportions of total traffic, by dividing each count by the total number of boardings at the stop (Figure \ref{ni39IBo_hP}). Peaks at three different time periods are now clearly visible.

%% stops: bo H

\begin{figure}[h]
\centering
\includegraphics[width=\widthcoef \textwidth]{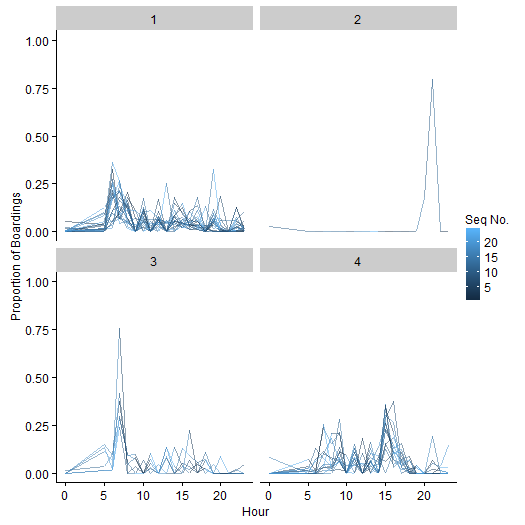}
\caption{Results of k-means clustering with Euclidean distance on the diurnal curves of proportional boardings.}
\label{ni39IBo_c50_hP_kmu4}
\end{figure}

These diurnal curves may simply be treated as vectors of length 24, and clustered with a variety of methods. Two approaches are shown here, each applied to the vector of proportion of boardings at each stop; for stability, the analysis includes only stops with at least 50 total boardings during the study period. The first approach uses the $L_2$ norm (Euclidean distance) to determine the distance between observations, and perform k-means clustering. Here, we set $k$ equal to 4; in practice, the number of clusters would be determined by how many separate policies transportation practitioners wished to develop. The resulting clusters, shown in Figure \ref{ni39IBo_c50_hP_kmu4}, appear to be dominated by rider behavior during two peak periods. Note that one cluster is devoted to a single stop, which features a peak of large absolute value at an unusual time.

%% 4 cl: kmu

\begin{figure}[h]
\centering
\includegraphics[width=\widthcoef \textwidth]{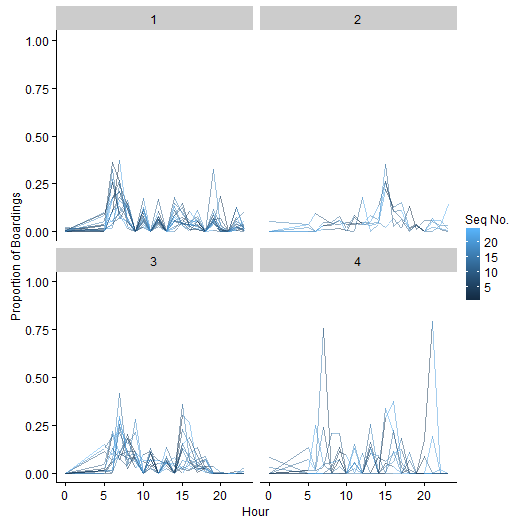}
\caption{Results of k-medoids clustering with band distance on the diurnal curves of proportional boardings.}
\label{ni39IBo_c50_hP_bd4}
\end{figure}

An alternative method is to use the band distance introduced in \cite{tupper}. This measure examines the dissimilarity of the rider curves at each pair of stops relative to the overall body of data, from all stops. Here it is paired with k-medoids clustering, with $k$ euqal to 4. In these results (Figure \ref{ni39IBo_c50_hP_bd4}), the single stop isolated under Euclidean distance is merged with other stops that seem to exhibit generally unusual behavior. The three other clusters appear to reflect different curve shapes, with one peak for observations in cluster 2, two peaks in cluster 3, and an early peak combined with sustained late activity in cluster 1.

%% 4 cl: bd

Both of these measures are based on the usage patterns of each bus stop; but stops also have inherent characteristics, based on their physical and network location, that can be used to determine similarity. While the dataset does not include information on the kinds of ``context" features used in \cite{zemp} and others, which relate to local land usage and demographic data, there is some spatial persistence in such features. Stops' physical distance from one another, then, may reflect these features to some degree. The physical distance between stops can be measured most straightforwardly as geographic Euclidean distance, based on the longitude and latitude of each stop.

Movement in cities, however, rarely follows Euclidean shortest paths; the effective distance between two bus stops depends on the road and transportation network between them. This idea has appeared previously in the planning literature: for example, \cite{lu} use this kind of network distance in determining distance-based weightings of local features, in the context of housing prices. Although this dataset does not have complete network information, it provides a useful proxy: the distance between two stops in terms of the bus route itself. This route distance between locations can be measured in two ways. The simplest method is to calculate how many stops apart the locations are, based on where each stop occurs in the route (called the ``global sequence number"). The dataset also records the cumulative distance that the bus has traveled when it reaches each doors-open event, so the difference in these cumulative values at two stops may also serve as the intervening travel distance. 

Because of the multiple variations of route 39, some of which include loops, these location-based distances are not entirely well behaved. Since data are aggregated across variations, it is necessary to choose only one location-based value (whether cumulative travel distance or sequential number along the route) for each stop, even though these values may in fact differ across trips, further obscuring the situation. Nevertheless, these metrics offer a different perspective on stop similarity and dissimilarity, based on fixed characteristics of each location rather than on the dynamics of its current usage.

\begin{figure*}[!h]
\centering
\begin{subfigure}{.3\textwidth}
\includegraphics[width= \textwidth]{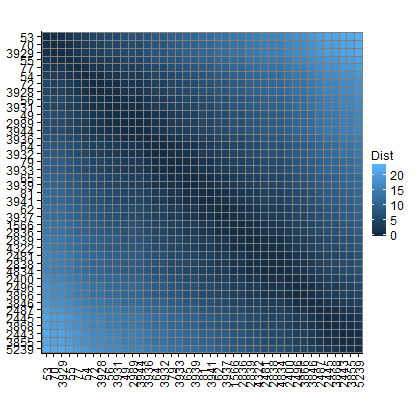}
\caption{Difference in global \\sequence number.}
\label{stopdist_gs_01}
\end{subfigure} \quad
\begin{subfigure}{.3\textwidth}
\includegraphics[width= \textwidth]{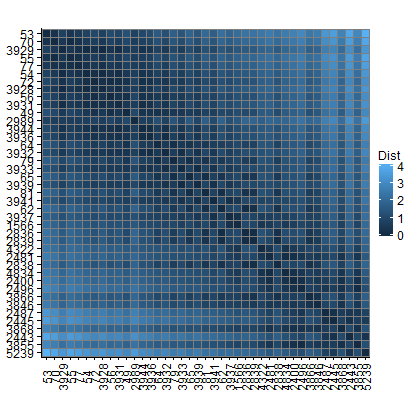}
\caption{Euclidean distance between geographic location.}
\label{stopdist_ll_01}
\end{subfigure} \quad
\begin{subfigure}{.3\textwidth}
\includegraphics[width= \textwidth]{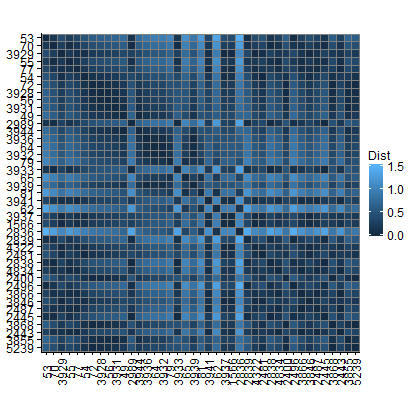}
\caption{Difference in cumulative travel distance.}
\label{stopdist_td_01}
\end{subfigure}

\begin{subfigure}{.3\textwidth}
\includegraphics[width= \textwidth]{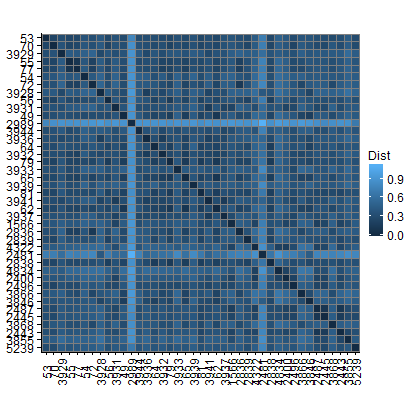}
\caption{Euclidean distance between boarding volume time curves.}
\label{stopdist_eu_01}
\end{subfigure} \quad
\begin{subfigure}{.3\textwidth}
\includegraphics[width= \textwidth]{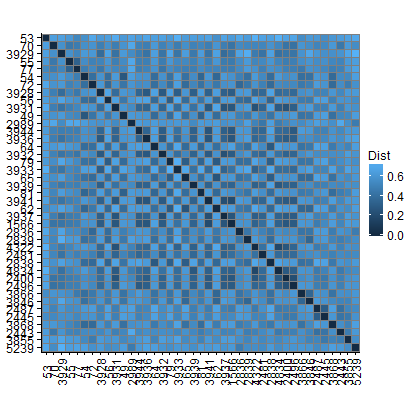}
\caption{Band distance between boarding volume time curves.}
\label{stopdist_bd_01}
\end{subfigure}
\caption{Matrices of pairwise distances between stops according to several metrics. Stops are ordered by global sequence number, proceeding top to bottom and left to right. Darker cells indicate relatively small distances between stops, lighter cells relatively large distances.}
\label{stopdists}
\end{figure*}

To visualize the differences between these metrics, it is possible to plot the pairwise distance matrices between stops according to each one (Figure \ref{stopdists}). In these plots, each row and column corresponds to a stop, ordered by the global sequence number of each stop. Darker cells indicate lower distance, and thus greater similarity, between stops; lighter cells indicate greater distances. Viewed as a whole, the distance matrices can reveal groups of stops with similar behavior, as well as individual stops that are distant from the rest. 

%% gSeq dist mat

The distance matrix for sequence number (Figure \ref{stopdist_gs_01}) is appropriately simple. Geographic Euclidean distance (Figure \ref{stopdist_ll_01}) shows strong agreement. Note that some pairs of stops shown in adjacent rows may have the same sequence number but occur on different variations, so that there is some (usually small) geographic distance between them. 

There are more notable differences when using cumulative travel distance (Figure \ref{stopdist_td_01}). This may be a database artifact, or a result of the loops and variations in the route; stop 2836, for example, occurs on two different route variations at very different points in the trip, so assigning it a single travel distance value is inevitably misleading for some trips.

Less straightforward behavior appears when stop dissimilarity is calculated based on usage patterns rather than location information. The distance matrix found using Euclidean distance on ridership time curves (Figure \ref{stopdist_eu_01}) reflects the clustering results above: the dominant feature is a single stop, 2989, found to be drastically different from the others. 

Using the band distance on ridership curves, in contrast, presents a more balanced picture (Figure \ref{stopdist_bd_01}). While some stops (including 2989) remain outliers dissimilar from all other stops, groups of noticeably similar stops appear as well. 

\begin{figure}[h]
\centering
\begin{subfigure}{.45\textwidth}
\includegraphics[width= \textwidth]{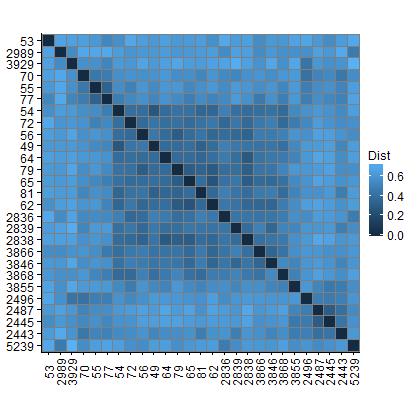}
\caption{Band distance.}
\label{stopdist_bd_01_v10}
\end{subfigure} \quad
\begin{subfigure}{.45\textwidth}
\includegraphics[width= \textwidth]{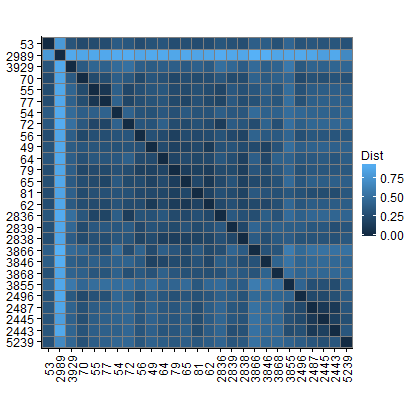}
\caption{Euclidean distance.}
\label{stopdist_eu_01_v10}
\end{subfigure}
\caption{Pairwise distances between stops according to distance between boarding volume time curves. Arranged by stops' order of incidence on variation 10.}
\label{stopdists_v10}
\end{figure}

It is interesting to plot a permuted version of this matrix, ordering the rows by stops' sequence along a single route variation (Figure \ref{stopdists_v10}). With this arrangement, it is clear that the band distance finds similarities between the ridership patterns of several groups of adjacent stops, reflected by lower distance values near the diagonal. For example, while the Transit Center (the last stop) remains distinct from all other stops, the nearby downtown stops within Rochester's inner road loop behave similarly to one another. Likewise, a large group of stops in the center of the trip, where the bus is traveling along major arterial streets, show related ridership behavior. The corresponding permutation of the Euclidean distance matrix is less enlightening; there is some agreement, but the overall picture is obscured by one or two extreme outliers.

%% band dist mat varn 10

\begin{figure}[H]
\centering
\includegraphics[width=\widthcoef \textwidth]{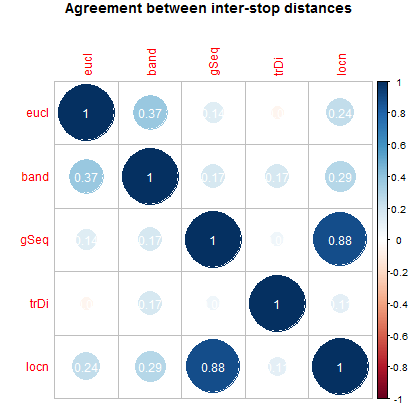}
\caption{Correlation (by Spearman's $\rho$) between the five distance measures: Euclidean and band distance on ridership curves, stop sequence number, cumulative travel distance, and geographic location.}
\label{cor_5dists_01}
\end{figure}

The natural next step, given multiple methods of determining dissimilarities between stops, is to examine how extensively they agree. A Pearson correlation is inappropriate here, since it would require assumptions about the distribution of distances under each metric. A better option is a nonparametric approach, first ranking pairs of stops from most to least distant according to each metric, then calculating Spearman's $\rho$. The resulting nonparametric correlations are shown in figure \ref{cor_5dists_01}.

%% dist corrplot

The two distances based on ridership patterns, using Euclidean and band distance (here labeled ``eucl" and ``band" respectively), show some agreement. Distances based on geographic location (``locn") and global sequential number along the route (``gSeq") strongly agree, and geographic location also shows some association with ridership-based measures of similarity. Using cumulative travel distance (``trDi"), however, gives results unrelated to any other metric, as was suggested by the distance matrix above. Future work will examine whether it is helpful to use a more sophisticated calculation of route-specific distances like travel distance and sequence number, by incorporating the differences between route variations.

%%%%%%%%
\section{Conclusion}
\label{sec:xer_conclu}
%%%%%%%%

This paper has presented an exploratory analysis of ridership data from the Rochester, NY bus system, toward characterizating system behavior and clustering individual stops into groups of interest to planners and policymakers. The results have demonstrated diurnal and weekly patterns, the skewed distribution of ridership volume across stops, and the relationship between ridership patterns and the system's orientation toward the city center and specifically the Transit Center. The analysis has also used several different metrics to examine the similarity and dissimilarity between stops, and shown an initial clustering of stops based on some of these measures.

Methodologically, there are many refinements to be made. Chief among them is a more informed selection of $k$, the number of clusters, using BIC or another measure of cluster viability. In practice, the number of clusters is limited by practical utility---transit planners and policymakers are unlikely to adapt programs to more than a handful of stop types---but previous work on clustering rail stations has found that the optimal number of clusters is fairly small, around five to ten (as in \cite{chen}, \cite{kim}, \cite{zemp}, \cite{higgins}). It is reasonable to assume that bus stops can also be placed into a fairly small number of classes.

A full characterization and classification of bus stops also requires more data as input. In terms of usage, this analysis has examined only boarding patterns on inbound stops, on weekdays, for illustration purposes. Alighting patterns, however, are also an important part of stop usage. These patterns are temporally distinct from boarding patterns, and it is possible that for some stops the two are not even in balance, as \cite{liu} found with taxi ridership. Weekend usage and seasonal changes in pattern (for example, during the summer tourist season or when schools are in session) also differentiate stops. Some stops, in addition, lie on multiple routes, and it would be necessary to combine data from all of these routes to see the overall usage pattern of the stop. The total amount of ridership information may become considerable, especially when considering many locations, and may require big-data methods for processing or reducing the dimension of the data.

This paper has addressed usage-based and location-based characterization of stops separately. Previous work (such as \cite{kim}, \cite{chen}, and \cite{zemp}) has touched on the relationship between these types of features, by using one type for classification and examining the association between cluster membership and the other feature type. But it would be possible to combine both types of features as input in future work, along with additional usage-related information like absolute level and variance. Information about local ``activity," as \cite{bertolini} puts it, and land use may offer a valuable perspective on stop types, especially in the absence of origin-destination pairs and demographic data on riders. Network-related characteristics---the ``node" features of \cite{bertolini} and \cite{reusser}, and measures of transit accessibility like those discussed in \cite{fu} and \cite{boisjoly}---may also come into play.

Combining these types of data leads to its own methodological questions, and points to the potential usefulness of the band distance. Euclidean distance is sensitive to the relative scaling of each dimension, and does not handle skewed data well; as discussed in \cite{tupper}, $L_p$-norm distances also become less informative in high dimensions. The band distance, by contrast, uses only the ordering of the observations on each dimension, and so could accommodate features that are scaled differently, skewed, or even not strictly numeric (such as ordinal estimates of non-quantitative factors like those discussed in \cite{zemp}).

Finally, depending on the goal, a purely stop-level analysis may not suffice. The work on classifying transit stations cited above has focused on classifying stops independently of one another, a reasonable approach for rail stations located a considerable distance apart; but bus stops, sited much more densely, may need to be considered in groups. While some physical facilities at a particular bus stop affect only riders boarding or exiting at that specific stop, other features of the area---walkability, bicycle facilities, intermodal connections, safety, and so on---are shared with nearby stops. The most obvious example is the paired stop, where a stop is located on each side of a street to serve a bus line running in two directions; but there are also groups of stops that share a local environment, like stops at different corners of an intersection that serve routes passing through on different streets. By combining data from multiple nearby stops into more general spatial nodes, an analysis could aid in planning not only individual stop facilities but the surrounding environment.

\section*{Acknowledgment}

Support was provided from a Xerox PARC Faculty Research Award and NSF Grant DMS-1455172.

\bibliographystyle{IEEEtran}
\bibliography{Tupper_et_al_Classification_arXiv}

\end{document}